\documentclass{elsart}

\usepackage{epsfig}

\usepackage{amssymb}

\begin{document}
\begin{frontmatter}

\title{Temperature dependences of the upper critical field and the Ginzburg-Landau parameter of Li$_2$Pd$_3$B from magnetization measurements}

\author[label1,label2]{I. L. Landau}
\author[label1]{R. Khasanov}
\author[label3]{K. Togano}
\author[label1]{H. Keller}
 \address[label1]{Physik-Institut der Universit\"at Z\"urich, Winterthurerstrasse 190, CH-8057 Z\"urich, Switzerland}
 \address[label2]{Institute for Physical Problems, 117334 Moscow, Russia}
  \address[label3]{National Institute for Materials Science 1-2-1, Sengen, Tsukuba, Ibaraki 305-0047, Japan}

%\author{}

%\address{}

\begin{abstract}
	
We present temperature dependences of the upper critical magnetic field $H_{c2}$ and the Ginzburg-Landau parameter $\kappa$ for a ternary boride superconductor Li$_2$Pd$_3$B obtained from magnetization measurements. A specially developed scaling approach was used for the data analysis. The resulting $H_{c2}(T)$ curve turns out to be surprisingly close to predictions of the BCS theory. The magnetic field penetration depth $\lambda$, evaluated in this work, is in excellent agreement with recent muon-spin-rotation experiments. We consider this agreement as an important proof of the validity of our approach.
\end{abstract}

\end{frontmatter}

%main text

Magnetization $(M)$ measurements in the mixed state of type-II superconductors are traditionally used for evaluation of the upper critical field $H_{c2}$. If magnetic fields $H$ higher than $H_{c2}$ are available, the upper critical field is usually detirmined as a value of $H$, at which diamagnetic moment of the sample vanishes. In real experimental situations, however, sample non-unformities and pining effects distort magnetization curves, making the choice of $H_{c2}$ ambiguous. In this work, we propose and use a scaling approach to the analysis of magnetization data. As will be shown, this approach allows to obtain  reliable $H_{c2}(T)$ results even for samples of moderate quality with considerable pinning effects. As an example, we consider magnetization data collected on a recently discovered ternary boride superconductor Li$_2$Pd$_3$B.

The idea of our analysis is simple. In magnetic fields $H$ much higher than the lower critical field $H_{c1}$, the equilibrium magnetic susceptibility in the mixed state of type-II superonductors may be written as
%%%%
\begin{equation}
\chi(H,T) = f_0(H/H_{c2})/\kappa^2,
\end{equation}
%%%%
where the function $f_0$ depends only on the ratio $H/H_{c2}(T)$, i.e., the temperature dependence of  $\chi$ is determined by temperature variations of $H_{c2}$ and $\kappa$. The relation expressed by Eq. (1) is rather general. Although it follows directly from the conventional Ginzburg-Landau theory of the mixed state, this equation is not limited to conventional superconductors only (see \cite{lokappa} for details). 

Multiplying both parts of Eq. (1) by $H/H_{c2}$ and introducing the universal magnetization function $m_0 = Hf_0(H/H_{c2})/H_{c2}$, we rewrite Eq. (1) as
%%%%
\begin{equation}
m_0(H/H_{c2}) = M(H,T)\kappa^2(T)/H_{c2}(T).
\end{equation}
%%%%
$m_0(H/H_{c2})$ depends neither on temperature nor on the absolute value of $\kappa$. This function may depend, however, on the symmetry of the order parameter or the sample geometry, i.e., it is not necessary the same for different superconductors. For conventional superconductors with zero demagnetizing factor $n$, $m_0(H/H_{c2})$ can be obtained, for instance, from numerical calculations of Brandt \cite{brandt}. Although  Eq. (2) is formally applicable only if $H \gg H_{c1}$, the actual limitations are not that strict.  As may be seen in Fig. 3 of Ref. \cite{lo-comm}, even for $\kappa$ as low as 5,  Eq. (2) represents a good approximation down to magnetic fields $H \sim 0.1H_{c2} \approx 2H_{c1}$.

Eq. (2) in a slightly different form was already used for the analysis of magnetization data collected on high-$T_c$ compounds \cite{lokappa,lo1}. Here we slightly modify that procedure  in oder to adjust it for materials with sufficiently low values of $H_{c2}$. In this case, the normal state paramagnetism may straightforwardly be evaluated from magnetization measurements above $H_{c2}$ and there is no need to introduce an additional adjustable parameter as it was done in \cite{lokappa,lo1}. 

\section{EXPERIMENTAL DETAILS}

The details of the sample preparation procedure can be found elsewhere \cite{togano}. The deviation of Li concentration from the stoichiometry was less than 1\% for this particular sample. A small (31 mg) part of the original (300mg) sample was used for magnetization measurements.\footnote{The original sample (300 mg) was used for muon-spin-rotation measurements in \cite{khas}} The sample had a platelike shape and it was placed with its longest side oriented along an applied magnetic field. The demagnetizing factor of the sample $n = 0.12$ was estimated by its magnetic susceptibility in the Meissner state. A commercial (Quantum Design) SQUID magnetometer was used for measurements.

\section{EXPERIMENTAL RESULTS AND ANALYSIS}

%%%%%%%%%%
\begin{figure}[h]
 \begin{center}
  \epsfxsize=0.8\columnwidth \epsfbox {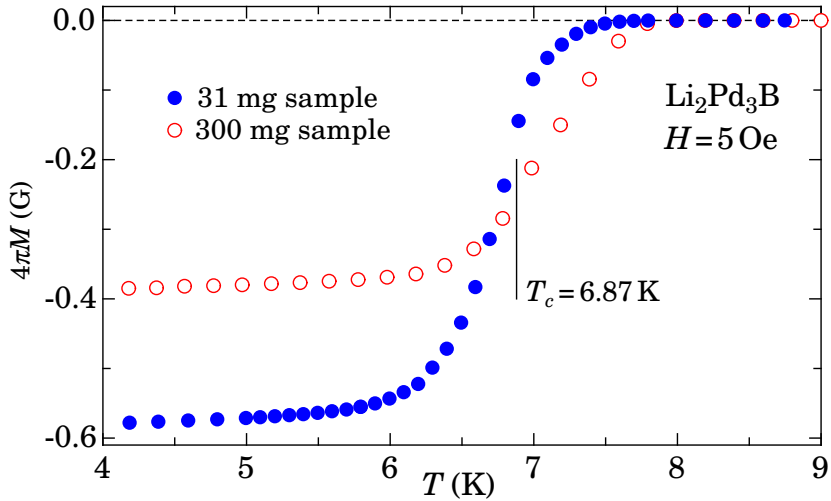}
  \caption{The field cooled magnetization measured in $H = 5$ Oe. The value of $T_c = 6.87$ K is estimated by the extrapolation of the $H_{c2}(T)$ curve to $H_{c2} = 0$, as will be explained in the text.}
 \end{center}
\end{figure}
%%%%%%%%%%%

Fig. 1 shows the field-cooled $M(T)$ curve measured in a low applied field of 5 Oe. As may be seen, the superconducting transition is rather broad ($\sim10\%$ of $T_c$), implying that the sample is not particular uniform. We also note that $4\pi M(T\rightarrow 0) \approx 0.1H \ll H$, i.e., pinning effects are strong. A similar curve for the original sample is also shown.$^1$ Comparing these two curves, we may conclude that the smaller sample has substantially narrower transition and considerably weaker pinning. We also note that the smaller sample, studied in this work, has somewhat lower critical temperature. 

Isothermal magnetization curves measured in increasing (zero-field-cooled) and decreasing fields at $T =4$ K are shown in the mainframe of Fig. 2(a). The $M(H)$ curve measured above $T_c$ is shown in the inset. As may be seen, a normal-state paramagnetic contribution to $M$ is substantial and it is not exactly linear on $H$. Nevertheless, because magnetization measurements at each temperature were extended to magnetic fields well above $H_{c2}$, the normal-state magnetization curves $M_n(H)$, measured above $H_{c2}$, could reliably be extrapolated to $H < H_{c2}$, as it is indicated by the solid line in Fig. 2(a). 
%%%%%%%%%%
\begin{figure}[t]
 \begin{center}
  \epsfxsize=0.8\columnwidth \epsfbox {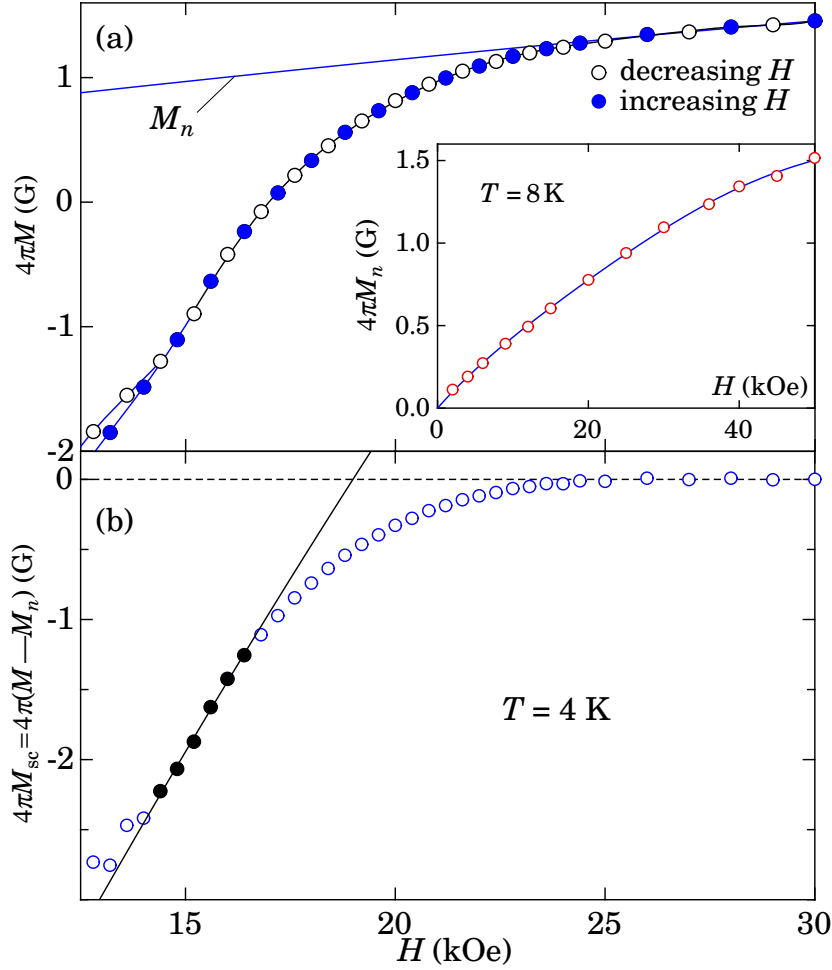}
  \caption{ (a) $M(H)$ curves below (4K) and above $T_c$ are shown in the mainframe and the inset, respectively. Extrapolation of $M_n$ to $H < H_{c2}$ is shown by the solid line.  (b) The diamagnetic response of the mixed state. The solid line is the best fit of theoretical calculations of Brandt \cite{brandt} to experimental data. The data points, which were used for fitting, are shown by black symbols.}
 \end{center}
\end{figure}
%%%%%%%%%%%

The difference $M_{sc} = M - M_n$ (diamagnetic response of the mixed state) is plotted in Fig. 2(b). In magnetic fields $H \gtrsim 14$ kOe, $M_{sc}(H)$ is reversible, i.e., represents  the equilibrium magnetization of the sample. The rounding of the experimental  $M_{sc}(H)$ curve in higher magnetic fields, which may clearly be seen in Fig. 2, is due to sample non-uniformity and is expected in this kind of samples. In spite of this rounding, there is a magnetic field range ($14 \le H \le 17$ kOe), in which $M_{sc}$ is a practically linear function of $H$, as it follows from the Abrikosov theory of the mixed state \cite{abr}. In the following we assume that this is the behavior of the main part of the sample. Using Eq. (2), the theoretical $m_0(H/H_{c2})$ curve, calculated in \cite{lo-comm}, can be fitted to experimental data with two fit-parameters $H_{c2}$ and $\kappa$.\footnote{In the evaluation of $\kappa$, we assume that the total volume $V_0$ of the sample contributes to $dM/dH$. In nonuniform samples this is not correct. The effective volume $V_{eff}$, which should be used in calculations, is smaller than $V_0$. This means that our estimation correspond to $\kappa_{eff} = \kappa \sqrt{V_0/V_{eff}} > \kappa$. We also note that one should expect a decrease of $V_{eff}$ with increasing temperature, which may lead to some distortion of the $\kappa(T)$ curve. However, because $\kappa$ is proportional to $\sqrt{V}$, these effects are not expected to be significant. } This kind of fitting was applied to magnetization curves measured in the temperature range 1.8 K$\le T\le 5.6$ K. The resulting temperature dependences of $H_{c2}$ and $\kappa$ are plotted in Fig. 3 by open circles. Because of stronger rounding of the magnetization curves at higher temperatures, this kind of fitting could only be used at $T < 5$ K. At  higher temperatures, as may be seen in Fig. 3, the values of $\kappa$ are unreliable. 

Both $H_{c2}(T)$ and $\kappa(T)$ agree rather well with theoretical calculations by Helfand and Werthamer (HW)  \cite{wert}.  This is why in the following analysis we  shall use $\kappa(T)$ as it is provided by this theory (the solid line in Fig. 3(b)). In this case, we have only one adjustable parameter $H_{c2}$ in Eq. (2), which greatly facilitates the fitting procedure and allows to extend the $H_{c2}(T)$ curve to higher temperatures. In all cases, we used for the analysis the steepest reversible part of the $M(H)$ curves, as it is illustrated in Figs. 2 (b) and 4(a). This kind of fit worked quite well from the lowest available temperature of 1.8 K up to $T \approx 6.2$ K. At higher temperatures, no reasonable fitting to experimental data points could be achieved (see Fig. 4(b)). This breakdown is the obvious consequence of the fact that $T_c$ is not uniform across the sample volume. The rounded part of the transition becomes broader with increasing temperature and at temperatures close to $T_c$ no linear part of the $M(H)$ curve can be identified (see Fig. 4(b)). 
%%%%%%%%%%
\begin{figure}[t]
 \begin{center}
  \epsfxsize=0.8\columnwidth \epsfbox {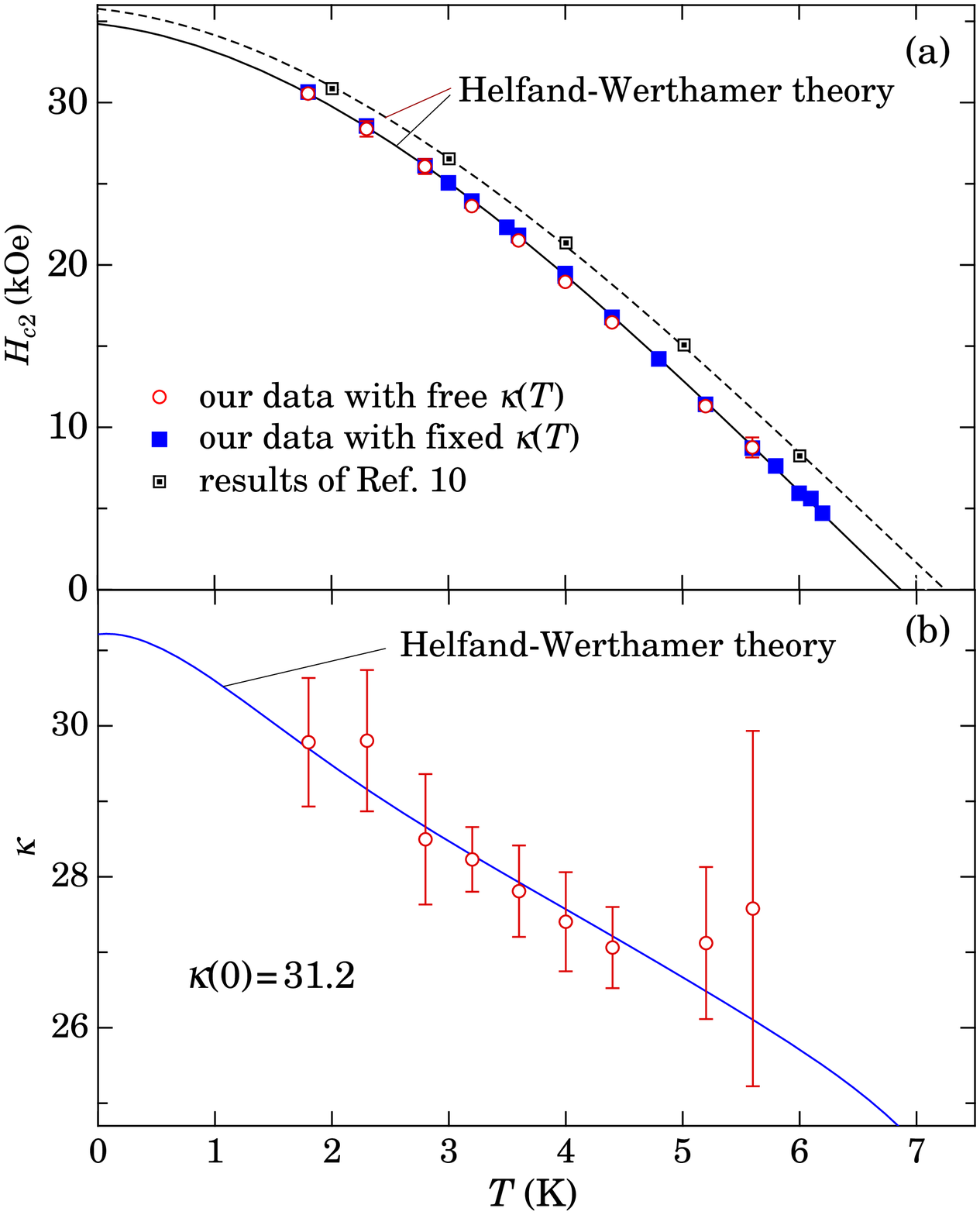}
  \caption{$H_{c2}(T)$ and $\kappa(T)$ evaluated by fitting of magnetization data to the theoretical $m_0(H/H_{c2})$ curve with two fit-parameters $H_{c2}$ and $\kappa$. The solid lines are the curves corresponding the Helfand-Werthamer theory \cite{wert}}. 
 \end{center}
\end{figure}
%%%%%%%%%%%
%%%%%%%%%%
\begin{figure}[b]
 \begin{center}
  \epsfxsize=0.8\columnwidth \epsfbox {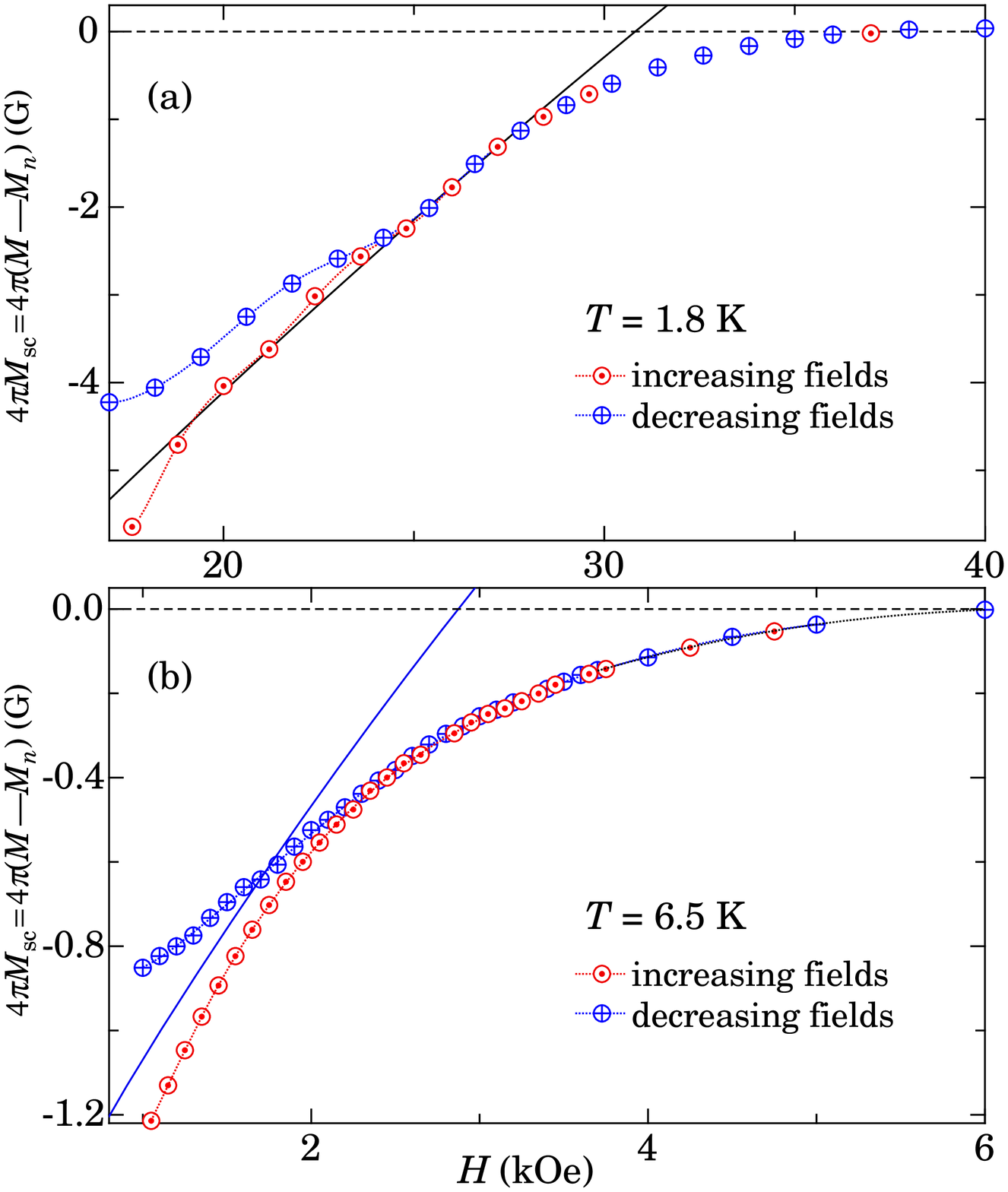}
  \caption{ (a) Magnetization curves. (a) $T=1.8$ K. The solid line represents theoretical $M(H)$ dependence fitted to experimental data as explained in the text. (b) $T=6.5$ K. As may be seen, the theoretical curve (the solid line) cannot be fitted to experimental data.}
 \end{center}
\end{figure}
%%%%%%%%%%%

The resulting $H_{c2}(T)$ dependence is shown in Fig. 3(a) by closed squares. It is in perfect agreement with the theory in the entire available temperature range. By  fitting of the theoretical $H_{c2}(T/T_c)$ curve to experimental data, we obtain $T_c = 6.87$ K and $H_{c2}(T=0) = 34.8$ kOe. The value of $T_c$, estimated in such a way, is consistent with low-field magnetization measurements (see Fig. 1). As has already been pointed out, the $H_{c2}(T)$ curve is evaluated quite reliably, while the values of $\kappa$, presented in Fig. 3(b), are overestimated.$^2$

In Ref. \cite{badica}, $H_{c2}$ was evaluated as a point of the onset of the diamagnetism in $M(H)$ and $M(T)$ measurements. This is the reason that the value of $H_{c2}(0)$, evaluated  in \cite{badica}, is higher than our result. We argue that the results of \cite{badica} do not represent the upper critical field as it is defined in the Ginzburg-Landau theory. 

We also show in Fig. 3(a) the $H_{c2}(T)$ curve for Li$_2$Pd$_3$B obtained from magnetization measurements by Doria et al. \cite{doria}. As may be seen, these data can also be very well fitted with the HW theory \cite{wert}. The resulting value of $H_{c2}(0) = 35.7$ kOe practically coincides with our result cited above. We note that in Ref. \cite{doria}, a very similar approach to the analysis of magnetization data was used. The only difference that the authors postulated that $\kappa$ is temperature independent. We argue, however,  that some decrease of $\kappa$ with increasing temperature may clearly be seen as an enhancement of slopes of scaled magnetization curves with increasing temperature (see Fig. 1 in \cite{doria}).  

%%%%%%%%%%
\begin{figure}[t]
 \begin{center}
  \epsfxsize=0.8\columnwidth \epsfbox {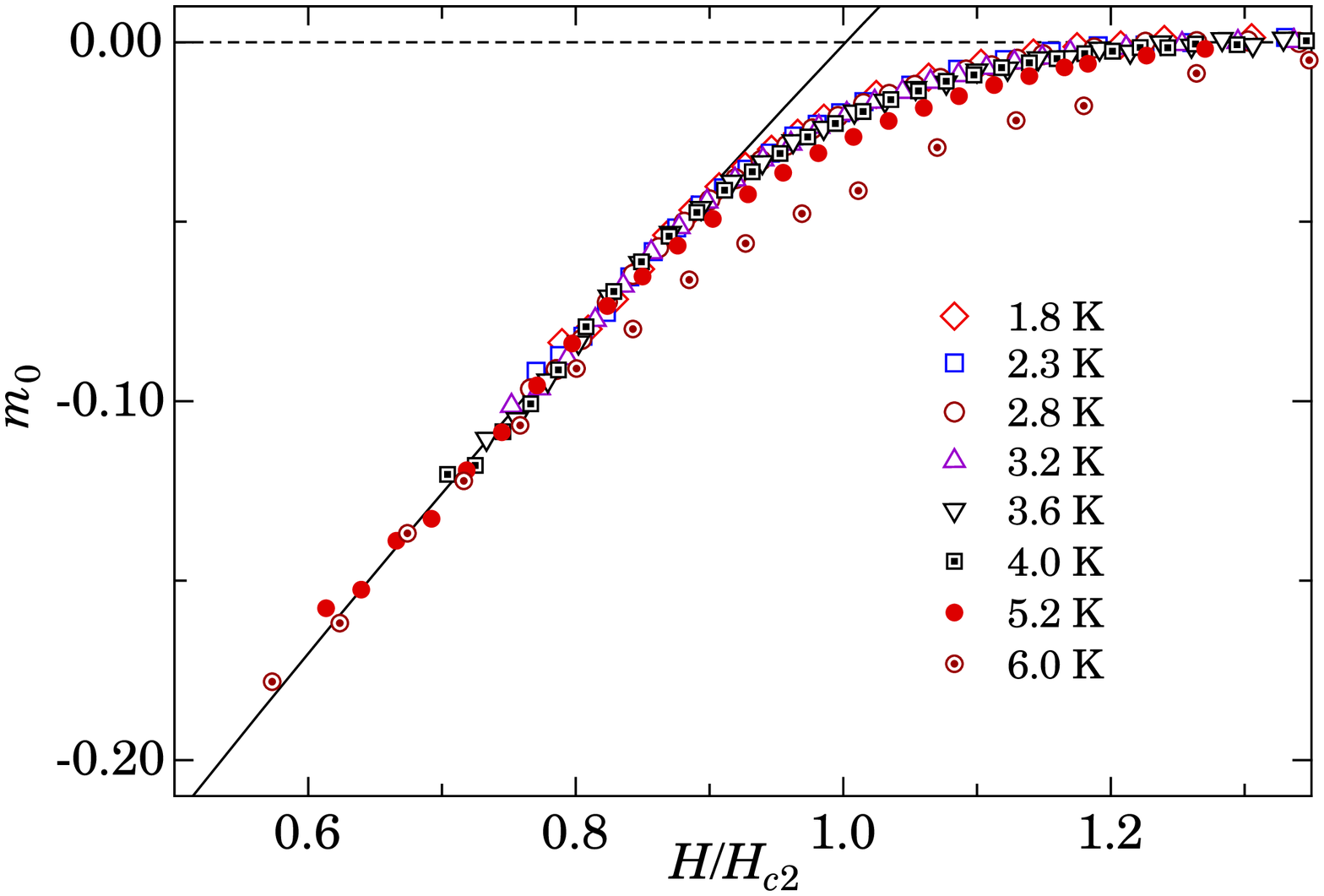}
  \caption{$m_0 = M(H,T)\kappa^2(T)/H_{c2}$ versus $H/H_{c2}$. Only reversible magnetization data are shown. The solid line represents the theoretical dependence according to \cite{brandt}.}
 \end{center}
\end{figure}
%%%%%%%%%%%
Using $H_{c2}(T)$ and $\kappa(T)$, as they are presented in Fig. 3(a), we can scale our $M(H)$ data according to Eq. (2) in order to obtain the $m_0(H/H_{c2})$ function. The results are presented in Fig. 5. In complete agreement with our assumptions, the data-points immediately above the irreversibility field collapse onto the theoretical curve. We also point out that at $T \le 4$ K all the data, including the rounded parts of the $M(H)$ curves, merge together. Noticeable deviations develop only at temperatures closer to $T_c$. A very similar behavior  was earlier observed in NbSe$_2$ samples \cite{NbSe}.

%%%%%%%%%%
\begin{figure}[b]
 \begin{center}
  \epsfxsize=0.8\columnwidth \epsfbox {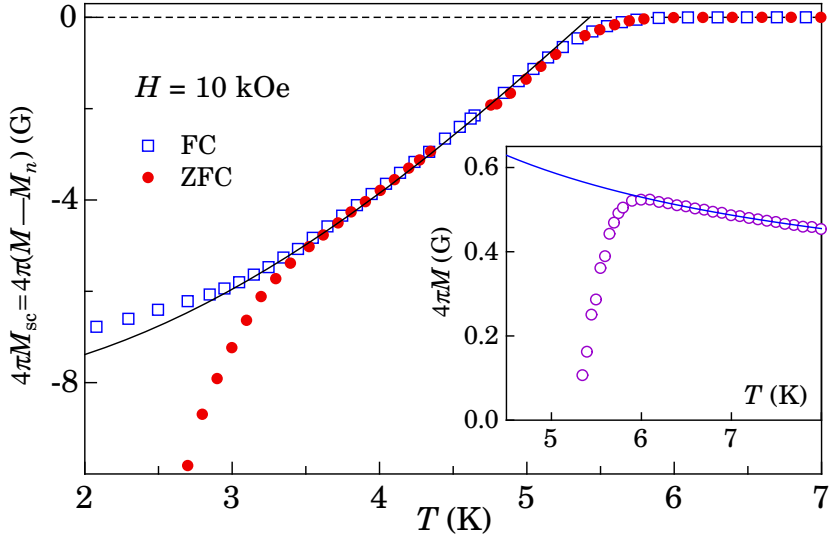}
  \caption{The magnetization $M(T)$ curve measured in a fixed magnetic field of 10 kOe. The solid line in the mainframe correspond to theoretical calculations for the Abrikosov vortex lattice \cite{brandt}. The inset shows original $M(T)$ data. The solid line represents the normal-state magnetization $M_n(T)$ extrapolated to $T< T_c$ assuming $M_n(T)\sim 1/T$.}
 \end{center}
\end{figure}
%%%%%%%%%%%
Finally, we present  the  $M(T)$ curve measured in a fixed magnetic field $H = 10$ kOe (Fig. 6). Because both $H_{c2}(T)$ and $\kappa (T)$ are already known, the theoretical curve can be calculated without any adjustable parameter. As may be seen in Fig. 6, agreement is rather good especially if errors in the extrapolation of $M_n(T)$ are taken into account. Here, at $T\gtrsim 5.2$ K, we can also see some rounding of the experimental $M(T)$ curve similar to that for isothermal $M(H)$ data (Figs. 2 and 4). Considering the results presented in Fig. 6, we want to underline that the theoretical $M(T)$ dependence is linear only in close vicinity of $T_c(H)$. This is why the linear extrapolation of experimental $M(T)$ curves usually results in overestimated values of $H_{c2}$ (see Fig. 6). Although this overestimation is not particularly strong, it may distort the resulting $H_{c2}(T)$ curve providing a positive curvature of $H_{c2}(T)$ at temperatures approaching $T_c$.

We also note a rather unusual onset of irreversibility, which may be seen in the isothermal $M(H)$ curves measured at temperatures well below $T_c$ (Figs. 2(b) and 4(a)). Indeed, both brunches (increasing and decreasing fields) of the curves deviate to the left from the corresponding solid lines. This effect is non-observable at temperatures close to $T_c$ (Fig. 4(b)). The origin of such a behavior is unclear. Similar observations in high-$T_c$ superconductors are usually explained by a vortex-lattice-melting transition. However, it is difficult to imagine that this kind of transitions can exist in samples of such poor quality. 

\section{DISCUSSION}

In the analysis above, we assumed that the temperature dependence of $\kappa$ follows the HW theory \cite{wert}. In complete agreement with this assumption, the slopes of the experimental $M(H)$ curves vary with temperature as $1/\kappa^2(T)$ in a rather extended range of temperatures $1.8 \le T \le 6.2$ K (see Figs. 2(b), 4(a) and 5). We consider this fact as evidence of the validity of our scaling approach. At temperatures $T > 6.2$ K, sample imperfections distort the magnetization curves in a way that they cannot and should not be used for evaluation of $H_{c2}(T)$.

In the HW theory \cite{wert}, $H_{c2}(T) = \sqrt{2}\kappa(T)H_c(T)$ ($H_c$ is the thermodynamic critical field) and both $H_c(T)$ and $\kappa(T)$ are calculated in the framework  of the weak coupling BCS theory. In principle, our result, presented in Fig. 5, may be considered as an indication that the zero-temperature energy gap $\Delta(0)$ is close to the corresponding weak coupling value $\Delta(0) = 3.52T_c$. However, in order to be sure that this is indeed the case, both $H_{c2}(T)$ and $\kappa(T)$ must coincide with theoretical predictions. In our case, as may be seen in Fig. 3(b), considerable pinning effects limit the accuracy to a degree that no unambiguous conclusion about temperature dependence of $\kappa$ can be made. In this sense, our results cannot be considered as contradicting to the value $\Delta(0) \approx 4T_c$ obtained for this compound in recent specific heat and muon-rotation experiments ($\mu$SR) \cite{khas,tak}.

Now we compare our results with those obtained in $\mu$SR experiments \cite{khas}. We remind that in this study we used a small part of  the sample studied in \cite{khas}.$^1$ The $\mu$SR experiments provide the variance of the magnetic induction in the sample, which is inversely proportional to $\lambda^4$. Fitting of of these results to the Ginzburg-Landau theory gives $H_{c2}^{(\mu)}(0) = 36.6$ kOe and $\lambda^{(\mu)}(0) = 2520$ \AA \cite{khas}. We use an index  $(\mu)$ to denote the results obtained in $\mu$SR experiments. 

Our value of $H_{c2}(0) = 34.8$ kOe practically coincides with $H_{c2}^{(\mu)}(0)$. Taking into account that the two values are obtained from completely different experimental data, this agreement should be considered as a rather convincing evidence that the Ginzburg-Landau theory may indeed be used for a quantitative description of the mixed state even at temperatures well below $T_c$ if the temperature dependence of $\kappa$ is taken into account.

Using $H_{c2}(T)$, we may calculate $\xi(T) = \sqrt{\Phi_0/2 \pi H_{c2}(T)}$ ($\Phi_0$ is the magnetic flux quantum). Extrapolation to $T=0$ gives $\xi(0) = 97$ \AA.  Because $\kappa(T)$ is also known, we can evaluate the magnetic field penetration depth as $\lambda(T) = \kappa(T)\xi(T)$. This gives $\lambda(0) = 3000$ \AA. This value is only 20 \% larger than  $\lambda^{(\mu)}(0)$. This is rather good agreement especially if we take into account that the samples are different. Furthermore, the two values of $\lambda(0)$ are not just close but the difference between them has a sign, which is expected. As was pointed out earlier,$^2$ our analysis gives slightly overestimated values of $\kappa$ and $\lambda$. At the same time, as it is discussed in \cite{khas}, in non-unuform samples, $\mu$SR experiments provide $\lambda$, which is smaller than its averaged value.

In conclusion, using a recently discovered Li$_2$Pd$_3$B superconductor, we demonstrate that the scaling analysis of the equilibrium magnetization may successfully be used in order to evaluate $H_{c2}(T)$, $\kappa(T)$, and $T_c$ from reversible magnetization data. It turned out that $H_{c2}(T)$ is very close to predictions of the HW theory clearly indicating that Li$_2$Pd$_3$B should be considered as a conventional single-gap superconductor. Our results agree well with recent muon experiments providing  $H_{c2}(0) = 34.8$ kOe, $\kappa(0) = 31.2$, $\xi(0) = 97$ \AA, and $\lambda(0) = 3000$ \AA.
%%%%%%%%%%%

\end{document}